\documentclass[11pt]{article}
\newcommand{\Comment}[1]{{}}
\usepackage{amsfonts,amsthm,amsmath,amssymb,slashed}
\usepackage[textwidth = 430 pt, textheight = 630 pt]{geometry}
\usepackage{color}

\Comment{\usepackage{color}
\definecolor{MyDarkBlue}{rgb}{0.15,0.15,0.45}
\usepackage[linktocpage=true]{hyperref}
\hypersetup{
colorlinks=true,
citecolor=MyDarkBlue,
linkcolor=MyDarkBlue,
urlcolor=MyDarkBlue,
pdfauthor={Horatiu Nastase and Carlos N\'{u}\~nez},
pdftitle={Deriving three-dimensional bosonization and the duality web},
pdfsubject={hep-th}
}

\usepackage[numbers,sort&compress]{natbib}
\usepackage{hypernat}}
\usepackage{graphicx}
\usepackage{cite}

\newcommand\ignore[1]{}
\def\one{{\,\hbox{1\kern-.8mm l}}}

\def\a{\alpha}

\def\d{\partial}

\def\dslash{\partial\!\!\!/}\def\Aslash{A\!\!\!\!/\,\,}

\newcommand{\Cset}{{\,\,{{{^{_{\pmb{\mid}}}}\kern-.45em{\mathrm C}}}}}

\newcommand{\be}{\begin{equation}}
\newcommand{\bea}{\begin{eqnarray}}

\newcommand{\ee}{\end{equation}}
\newcommand{\eea}{\end{eqnarray}}

\begin{document}

\renewcommand{\thefootnote}{\fnsymbol{footnote}}

\makeatletter
\@addtoreset{equation}{section}
\makeatother
\renewcommand{\theequation}{\thesection.\arabic{equation}}

\rightline{}
\rightline{}
%   \vspace{1.8truecm}

%\begin{flushright}
% preprint nrs.
%\end{flushright}

%\vspace{10pt}

%%%%%%%%%%%%%%%%%

\begin{center}
{\LARGE \bf{\sc Deriving  three-dimensional bosonization and the duality web}}
\end{center} 
 \vspace{1truecm}
\thispagestyle{empty} \centerline{
{\large \bf {\sc Horatiu Nastase${}^{a}$}}\footnote{E-mail address: \Comment{\href{mailto:nastase@ift.unesp.br}}{\tt nastase@ift.unesp.br}}
{\bf{\sc and}}
{\large \bf {\sc Carlos N\'{u}\~nez${}^{b}$}}\footnote{E-mail address: \Comment{\href{mailto:c.nunez@swansea.ac.uk}}{\tt c.nunez@swansea.ac.uk}}
                                                        }

\vspace{.5cm}

%\vspace{.3cm}

\centerline{{\it ${}^a$Instituto de F\'{i}sica Te\'{o}rica, UNESP-Universidade Estadual Paulista}} 
\centerline{{\it R. Dr. Bento T. Ferraz 271, Bl. II, Sao Paulo 01140-070, SP, Brazil}}
\vspace{.3cm}
\centerline{{\it ${}^b$Department of Physics, Swansea University,}}
\centerline{{\it Swansea SA2 8PP, United Kingdom}}

\vspace{1truecm}

%%%%%%%%%%%%%%%%%
\thispagestyle{empty}

\centerline{\sc Abstract}

\vspace{.4truecm}

\begin{center}
\begin{minipage}[c]{380pt}
{\noindent Recently, a duality web for three dimensional theories with Chern-Simons terms was proposed. This can be derived from a single 
bosonization type duality, for which various supporting arguments (but not a proof) were given. Here we explicitly derive this bosonization, in the Abelian case 
and for a particular regime of parameters. 
To do this, we use the  particle-vortex duality in combination with a Buscher-like duality (both considered in the regime of  low energies). As a corollary, 
Son's conjectured
duality is derived in  a somewhat singular limit of vanishing mass.
}
\end{minipage}
\end{center}

\vspace{.5cm}

\setcounter{page}{0}
\setcounter{tocdepth}{2}

\newpage

%\tableofcontents
\renewcommand{\thefootnote}{\arabic{footnote}}
\setcounter{footnote}{0}

\linespread{1.1}
\parskip 4pt

%{}~
%{}~

%---------------------------------------------------------

%%%%%%%%%%%%%%%%%%%%%%%%%%%%%%%%%%%%%%%%%%%%%%%%%%%%%%%%%%%%%%%%%%%%%%%%%%%%%%%%%%%%%%%%
\section{Introduction}
%%%%%%%%%%%%%%%%%%%%%%%%%%%%%%%%%%%%%%%%%%%%%%%%%%%%%%%%%%%%%%%%%%%%%%%%%%%%%%%%%%%%%%%%

Duality symmetries are powerful tools that serve to constrain and understand non-perturbative physics. In (2+1)-dimensions,  within the context of 
condensed matter systems, dualities  have received less attention in comparison to their (3+1)-dimensional counterparts, that naturally 
appear in  particle physics. Recently however, 
partly motivated by the desire to understand Son's conjecture
 \cite{Son:2015xqa} a web of dualities was proposed \cite{Murugan:2016zal,Seiberg:2016gmd,Karch:2016sxi}.
 
 In fact, D. T. Son has proposed a relation  between a massless Dirac 'fundamental' fermion and a 'composite' Dirac fermion coupled to a 
 gauge field with BF-dynamics
  \cite{Son:2015xqa}.  The 'fundamental' fermion is to be understood as a boundary mode in a topological insulator, while the 'composite' 
  one should be thought of as an effective description for a half-filled lowest Landau level  of a Fermi liquid  \cite{Son:2015xqa},\cite{Metlitski:2015eka}. 
  The whole idea is driven by the field theoretical descriptions of a (time reversal invariant) topological insulator
and a topological superconductor.

The web of dualities mentioned above relates various bosonic theories (for scalars and gauge fields) with  fermionic theories (coupled to a 
vector field), both  with Chern-Simons terms. All fields transform under a $U(1)$ gauge symmetry. Extensions, including to non-Abelian 
cases have been considered in \cite{Hsin:2016blu,Karch:2016aux,Aharony:2016jvv,Kachru:2016aon,Radicevic:2016wqn}.

%
 %relation between a massless Dirac fermion, understood as the boundary mode of a 
%topological insulator, and the composite fermion describing the low energy half-filled Landau level of a Fermi liquid \cite{Son:2015xqa}
%(see also \cite{Metlitski:2015eka}), and also guided by the field theory description of a 3+1 dimensional time reversal invariant 
%topological insulator \cite{Qi:2008ew} and of a topological superconductor \cite{Qi:2012cs}, a web of dualities was proposed 
%\cite{Murugan:2016zal,Seiberg:2016gmd,Karch:2016sxi}. The dualities relate various bosonic and fermionic theories with Chern-Simons terms, 
%and while the original one was Abelian, extensions including to non-Abelian cases were considered also, for instance 
%\cite{Hsin:2016blu,Karch:2016aux,Aharony:2016jvv,Kachru:2016aon,Radicevic:2016wqn}.

The web of dualities can be derived by assuming the validity of a  basic correspondence between a bosonic theory and a fermionic one, which
in the rest of this paper will be referred to as {\it three-dimensional bosonization}. 
These ideas were  considered and extended to the context of supersymmetric theories by Aharony 
\cite{Aharony:2015mjs}. 

On the other hand, an explicit mapping between a bosonic and a fermionic theory was presented around twenty years ago 
in \cite{Burgess:1993np},\cite{Burgess:1994tm,Fradkin:1994tt}. It is based on the realisation that two-dimensional bosonization can 
be viewed as a Buscher-like duality and the  extension of such procedure to three dimensions.
We will refer to it as the {\it Burgess-Quevedo map} (or BQ-map). See the paper  \cite{Schaposnik:1995np}  for a careful account of 
the idea and technical details of the  BQ-map.

Postulating the validity of the three-dimensional bosonization duality,  one can derive the (bosonic) particle-vortex duality, or the fermionic 
duality conjectured by Son. 
Repeated application leads to a full duality web. While this  basic three-dimensional bosonization was not proven,  evidence indicative 
of its correctness was presented in 
\cite{Seiberg:2016gmd,Karch:2016sxi} and subsequent papers, e.g. \cite{Filothodoros:2016txa}.

In this note, we will derive this basic three-dimensional bosonization conjecture as presented in \cite{Karch:2016sxi,Karch:2016aux}.
To do this, we will assume the validity of the particle-vortex duality and  combine it with the  Buscher-like  
BQ-map \cite{Burgess:1993np}, \cite{Burgess:1994tm},\cite{Fradkin:1994tt}. In fact, a  regime of sufficientely low energies,
 with special field configurations in the particle-vortex equivalence, together with a BQ-map improved by the presence of  point like vortices in the system, are instrumental to our derivation.
 
 Our approach will be phrased in  the framework of the path integral formalism, as  defined, for example,  in 
\cite{Murugan:2014sfa}, which is itself  based on the earlier work  \cite{Burgess:2000kj} (see also \cite{Ramos:2005yy,Ramos:2007hk} for an alternative 
viewpoint and \cite{Zee:2003mt} for the usual condensed matter formulation). The particle-vortex duality was  discussed in 
works on superconductivity \cite{Dasgupta:1981zz}, \cite{Marino:1987tk,Marino:1992uu}, and also in the contexts of anyon superconductivity and the fractional 
quantum Hall effect  \cite{Lee:1989fw}.

This work is organized as follows. In Section \ref{section2}, we summarize and streamline the background material needed 
for our purposes: the three-dimensional bosonization proposal, time reversal, Son's duality and the particle vortex duality. In Section \ref{section3} we 
derive the conjectured three dimensional bosonization, assuming the validity of the particle-vortex duality and the BQ-map.
Section \ref{conclusion} closes the paper with final conclusions.

%%%%%%%%%%%%%%%%%%%%%%%%%%%%%%%%%%%%%%%%%%%%%%%%%%%%%%%%%%%%%%%%%%%%%%%%%%%%%%%%%%%%%%%%
\section{Three-dimensional bosonization and the duality web}\label{section2}
%%%%%%%%%%%%%%%%%%%%%%%%%%%%%%%%%%%%%%%%%%%%%%%%%%%%%%%%%%%%%%%%%%%%%%%%%%%%%%%%%%%%%%%%

As a warm-up, in this section we will review how (part of) the Abelian duality web is derived. We will also discuss the action of time reversal on 
the different dualities and go over the derivation in \cite{Seiberg:2016gmd,Karch:2016sxi} of 
Son's conjectured relation \cite{Son:2015xqa}. Finally, the particle-vortex duality will be shown to arise from 
alternate integrations on a 'master' partition function that depends on both 'particle' and 'vortex' fields.

A main basic ingredient in this work is the three dimensional bosonization that we now review,  adopting  the notation
in \cite{Karch:2016sxi}. The partition functions for a complex scalar $\phi=\phi_0 e^{i\theta}$ coupled to a vector $A_\mu$ (adding 'flux'), and 
that for a Dirac fermion $\psi$ (both in the presence of a vectorial external source $S_\mu$) are
\bea
\tilde{Z}_{\rm scalar+flux}[S]&\equiv&\int {\cal D}\phi {\cal D}\phi^* {\cal D}A_\mu e^{iS_{\rm scalar}[\phi,A]+iS_{\rm CS}[A]+iS_{\rm BF}[A,S]}\cr
&=&\int {\cal D}\phi_0{\cal D}\theta {\cal D}A_\mu e^{iS_{\rm scalar}[\theta,A;\phi_0]-\frac{1}{2}\int d^3x(\d_\mu\phi_0)^2+iS_{\rm CS}[A]+iS_{\rm BF}[A,S]}\cr
Z_{\rm fermion}[S;m]&\equiv& \int {\cal D}\psi{\cal D}\bar\psi e^{i\int \bar\psi(\dslash+m +S)\psi}.
\eea
We have  denoted,
\bea
S_{\rm CS}[A]\equiv \frac{1}{4\pi}\int d^3x \epsilon^{\mu\nu\rho}A_\mu\d_\nu A_\rho,\;\;\;\;
S_{\rm BF}[A,S]\equiv  \frac{1}{2\pi}\int d^3x \epsilon^{\mu\nu\rho}A_\mu \d_\nu S_\rho.
\eea
The action for the complex scalar $\phi$ is defined and can be rewritten according to,
\bea
S_{\rm scalar}[\phi,A] \equiv -\frac{1}{2}\int d^3x |(\d_\mu -iA_\mu)\phi|^2 
\rightarrow S_{\rm scalar}[\theta,A;\phi_0] \equiv -\frac{1}{2}\int d^3x \phi_0^2(\d_\mu\theta+A_\mu)^2.\label{xxy}
\eea
Note that the scalar action $S_{\rm scalar}[\theta,A;\phi_0]$
in the last expression  of eq.(\ref{xxy}) appears for the case in which the modulus $\phi_0$ is constrained to be 
constant. Such an action is obtained from that of a complex scalar $\phi$ with a symmetry breaking Higgs-like potential, 
\bea
S_{\rm scalar}[\theta,A;\phi_0]&=&\lim_{\alpha\to\infty} S_{\rm scalar}[\theta,A;\phi_0]
-\int d^3x\left[ \frac{1}{2}(\d_\mu\phi_0)^2+\frac{\a}{2}(\phi_0^2-m)^2\right]\cr
&=&\lim_{\alpha\to\infty}S_{\rm scalar}[\phi,A]-\int d^3x \frac{\a}{2}(\phi_0^2-m)^2\;,
\label{Higgsdef}
\eea
with  the coupling $\a$ taken to be very large, $\a\rightarrow\infty$. Equivalently,  for low energies $E\ll \a$, the quantity $\phi_0$ takes a constant value. In most of the 
analysis below we will consider $\phi_0$ to be fixed, $\phi_0=\sqrt{m}$, and we will drop $\int {\cal D}\phi_0$ from the path integral.

Then, the basic three-dimensional bosonization  duality, relates a 
fermion coupled to a background vectorial current with a complex scalar plus flux, considered in general with a fluctuating $\phi_0$ (hence the tilde on $
Z_{\rm scalar+flux}$). More  explicitly, 
\be
Z_{\rm fermion}[S;m=0]e^{-\frac{i}{2}S_{\rm CS}[S]}=\tilde Z_{\rm scalar+flux}[S]\;.\label{bosonization}
\ee
In the paper \cite{Karch:2016aux},  the authors proposed a more general duality for the bosonization of  a massive fermion (of mass $m$). 
This extended relation 
that leads to a more general web of dualities reads
\bea
Z_{\rm fermion}[S;m]e^{-\frac{i}{2}S_{\rm CS}[S]}&=& Z_{\rm scalar+flux}[S].\label{bosonization2}\\
Z_{\rm scalar+flux}[S]&=&\!\!\!\!\!\!\!
\lim_{{\!\!\a\rightarrow \infty,E\ll \a}}\!\!\int\!\!\!{\cal D}A_\mu {\cal D}\phi_0{\cal D}\theta {\cal D}\sigma\cr
&&e^{\!iS_{\rm scalar }(\theta,A;\phi_0)\!+\!iS_{\rm CS}[A]\!+i\!S_{\rm BF}[A,S]\!-i\!\int\! d^3x\! \left[\frac{1}{2}(\d_\mu\phi_0)^2
+\sigma(\phi_0^2-m)+\frac{\sigma^2}{2\a}\right]}.\nonumber
\eea
In the case of vanishing mass ($m=0$), integrating out the non-dynamical field $\sigma$ we generate a potential $V=\phi_0^4/2\a$, which 
leads to the Wilson-Fisher fixed point at low energies. If $m>0$, integrating out $\sigma$ we get the Higgs-like potential in eq. (\ref{Higgsdef}). At 
small enough energies $E\ll m=\phi_0^2$, $E\ll \a$, the dynamical  field $\phi_0$ freezes-out, leaving us simply with 
$Z_{\rm scalar+flux}[S]$ on the right hand side (notice that the integration in $\phi_0$ is trivial, hence the absence of tilde in $Z_{\rm scalar+flux}[S]$). 
More explicitly, at low energies and after the constraint is imposed, we have
\bea
Z_{\rm scalar+flux}[S]=\int {\cal D}A_\mu {\cal D}\theta
e^{iS_{\rm scalar }(\theta,A;\phi_0)+iS_{\rm CS}[A] +i S_{\rm BF}[A,S]}.\label{papa}
\eea
In the following we will consider the situation in which the constraint $\phi_0^2=m$ is enforced by the integration over the field $\sigma$, in the limit of low energies. 
More precisely, we will probe the dynamics with energies that are very small compared to those set by the two relevant scales, $m$ and $\a$. 

{\bf Time-reversed relation}

Another ingredient needed to  prove different entries of  the duality web comes from  considering  the effect of time reversal on the system. 
Time reversal invariance
 leads to relations, which  change the sign of the Chern-Simons and BF terms. Indeed, we also have the duality,
\be
Z_{\rm fermion}[S]e^{+\frac{i}{2}S_{\rm CS}[S]}=\bar{\tilde{Z}}_{\rm scalar+flux}[S]\equiv 
\int {\cal D}\phi {\cal D}\phi^* {\cal D}A_\mu e^{iS_{\rm scalar}[\phi,A]-iS_{\rm CS}[A]-iS_{\rm BF}[A,S]}.\label{timerev}
\ee
The bosonic and fermionic particle-vortex dualities are obtained by applying and manipulating the three-dimensional bosonization 
relation in eq. (\ref{bosonization}), 
 and using then the time-reversed bosonization relation above. 

{\bf Son's duality from bosonization}

As an example, we derive Son's conjectured duality between a massless Dirac fermion $\psi$ coupled to an external field $S_\mu$ and a 
composite Dirac fermion
$\chi$, coupled to a dynamical field $A_\mu$, which itself couples to the external $S_\mu$ through a BF coupling, denoted BF-QED. 
In what follows, we summarise 
a derivation in \cite{Seiberg:2016gmd},\cite{Karch:2016sxi}. Indeed, the 
dynamics of the composite fermion $\chi$ and the vector $A_\mu$ is described by
\be
Z_{\rm BF-QED}[S;m]=\int {\cal D}A_\mu {\cal D}\chi{\cal D}\bar\chi e^{i\int \bar\chi(\dslash +m+\Aslash)\chi+\frac{i}{2}S_{\rm BF}[A,S]}\;.
\ee
Son conjectured a duality between the composite, low energy, massless BF-QED theory (set $m=0$ in the above $Z_{\rm BF-QED}$) 
and a massless  Dirac fermion theory, both coupled to an external source $S_\mu$,
\be
Z_{\rm BF-QED}[S]=Z_{\rm fermion}[S]\;.\label{Son}
\ee
To derive eq.(\ref{Son}), one starts from eq.(\ref{bosonization}), changing the notation as $S_\mu\rightarrow \bar A_\mu$, adds 
a BF term $\frac{i}{2}S_{\rm BF}
[\bar A,S]$ (where now $S_\mu$ is a new external field) on both sides. 
Takes the $e^{-\frac{i}{2}S_{\rm CS}[S]}$ to the other side, and then integrates over $\bar A_\mu$ (formerly, the external field). 
Then the left hand side becomes
$Z_{\rm BF-QED}[S]$, while the right hand side turns into
\be
\tilde{Z}_{\rm scalar+fluxes}[S]=\int {\cal D}\phi {\cal D}\phi^* {\cal D}A_\mu {\cal D}\bar A_\mu 
e^{iS_{\rm scalar}[\phi,A]+iS_{\rm CS}[A]+iS_{\rm BF}[A,\bar A] +\frac{i}{2}S_{\rm BF}[\bar A,S]+\frac{i}{2}S_{\rm CS}[\bar A]}.\label{mama}
\ee
Performing  the integration over $\bar A_\mu$ (which appears algebraically), we find the equation of motion $d\bar A=-(dS+2dA)$. Finally replacing 
$\bar{A}_\mu=-(S_\mu+2 A_\mu)$  back in the scalar partition function of eq.(\ref{mama}), we find
\be
Z_{\rm BF-QED}[S]=\int {\cal D}\phi {\cal D}\phi^*  {\cal D}A_\mu 
e^{iS_{\rm scalar}[\phi,A]-iS_{\rm CS}[A]-iS_{\rm BF}[A,S] -\frac{i}{2}S_{\rm CS}[S]}\;,
\ee
which because of eq.(\ref{timerev}) (the {\em time-reversed} form of the basic bosonization duality) 
equals $Z_{\rm fermion}[S]$. The final result is Son's relation in eq.(\ref{Son}). 

A new result can be obtained if we start from   the three-dimensional bosonization proposal in eq.(\ref{bosonization2}), follow exactly the same procedure
described above and derive a Son-like relation between a fundamental and a composite Dirac fermions, both with the same mass $m$,
\bea
Z_{\rm BF-QED}[S;m]= Z_{\rm fermion}[S;m].
\label{Son2}
\eea
In Section \ref{section3}, we will put this last correspondence on a firmer basis, by proving the equivalence in eqs.(\ref{bosonization2})-(\ref{papa}).
Let us now revisit another important duality.

{\bf Review of the particle-vortex duality }

Another ingredient needed in our derivation of Section \ref{section3},  is a specific form of a particle-vortex duality.
In the paper \cite{Murugan:2014sfa}, a transformation  was proposed that realizes a particle-vortex duality as an equivalence of partition functions.
Getting rid of some unnecessary (for our purposes) extra ingredients, the two partition functions that are shown to be equivalent are
\be
Z_{\rm particle}=\int {\cal D}\theta e^{iS}\equiv\int {\cal D}\theta \exp\left[-i\int d^3x \frac{1}{2}\left[(\d_\mu\phi_0)^2+\phi_0^2(\d_\mu\theta_{\rm smooth}
+\d_\mu\theta_{\rm vortex}+A_\mu)^2\right]\right]\;,
\label{particlez}\ee
and
\bea
& & Z_{\rm vortex}=\int{\cal D}\lambda_\mu e^{iS_{\rm dual}} \label{pvdual}\\ 
&=&\int {\cal D}\lambda_\mu \exp\left[-i\int d^3x\left[\frac{1}{2}(\d_\mu\phi_0)^2+\frac{1}{4(2\pi\phi_0)^2}f^{(\lambda)}_{\mu\nu} f^{(\lambda) \mu\nu}
\right.\right. \left.\left.+\frac{1}{2\pi}\epsilon^{\mu\nu\rho}\lambda_\mu\d_\nu A_\rho+ j^\mu_{\rm vortex}\lambda_\mu\right]\right]\;.\nonumber\eea
Let us clarify the different terms in these expressions.
The expression in eq.(\ref{particlez}) is written in terms of a dynamical field $\theta$ and two external ones $A_\mu$ and $\phi_0$.
We  shall separate $\theta$ into a dynamical smooth part $\theta_{\rm smooth}$ and a nondynamical vortex part $\theta_{\rm vortex}$
 that contains singularities
(at $r=r_v$), i.e. $\int_0^{2\pi} d\a \d_\a \theta_{\rm vortex}=2\pi N$, where $\a$ is the polar angle in 2 spatial dimensions, measured with respect to the 
positions $r=r_v$ of vortices. Thus the integral ${\cal D}\theta$ splits into $\int {\cal D}\theta_{\rm smooth}$ times a sum over the nontrivial 
vortex numbers $\sum_N$ for the various $\theta_{\rm vortex}$ sectors. More explicitly $\int {\cal D}\theta=\sum_{N}\int {\cal D}\theta_{\rm smooth}$.

On the other hand,  the partition function in eq.(\ref{pvdual}) is written in terms of a dynamical vector $\lambda_\mu$, with external sources $A_\mu$ and $\phi_0$. We have  defined
$f_{\mu\nu}^{(\lambda)}\equiv \d_\mu \lambda_\nu-\d_\nu \lambda_\mu$ and the vortex current,
\be
j^\mu_{\rm vortex}\equiv \frac{1}{2\pi}\epsilon^{\mu\nu\rho}\d_\nu\tau_{\rho, {\rm vortex}}
=\frac{1}{2\pi}\epsilon^{\mu\nu\rho}\d_\nu\d_\rho\theta_{\rm vortex}.\label{yyyy}
\ee
Let us now show the equivalence,
\bea
Z_{\rm particle}=Z_{\rm vortex}.\nonumber
\eea
In order to do this,  we use the usual trick of constructing a master partition function (dependent on two variables), that reduces either to 
$Z_{\rm particle}$, or the dual vortex one $Z_{\rm vortex}$, upon alternate integration-out of one or the other variable.

To construct such master path integral, first 
replace $\d_\mu\theta=\d_\mu \theta_{\rm smooth}+\d_\mu \theta_{\rm vortex}$ with an independent variable $\tau_\mu=\tau_{\mu,{\rm smooth}}
+\tau_{\mu,{\rm vortex}}$, and then impose the flatness of the smooth part's curvature by $\epsilon^{\mu\nu\rho}\d_\nu \tau_{\rho,{\rm smooth}}=0$, 
with  Lagrange multiplier $\lambda_\mu$. We then obtain the master partition function,
\bea
Z_{\rm master}&=&\int {\cal D}\tau_\mu{\cal D}\lambda_\mu e^{iS_{\rm master}}\cr
&\equiv&\int {\cal D}\tau_\mu{\cal D}\lambda_\mu\exp\left[i\int d^3x \left\{  -\frac{1}{2}(\d_\mu\phi_0)^2\right.\right.\cr
&&\left.\left.-\frac{1}{2}\phi_0^2(\tau_{\mu,{\rm smooth}}+
\tau_{\mu,{\rm vortex}}+A_\mu)^2+\frac{1}{2\pi}\epsilon^{\mu\nu\rho}\lambda_\mu\d_\nu \tau_{\rho,{\rm smooth}}\right\}\right]\;,
\label{masterxx}
\eea
where again $\int{\cal D}\tau$ is understood as $\sum_{N}\int {\cal D}\tau_{\mu,{\rm smooth}}$.

If we solve for the Lagrange multiplier $\lambda_\mu$ (and integrate it out), we obtain $\tau_{\mu,{\rm smooth}}=\d_\mu\theta_{\rm smooth}$.
Substituting it back into eq.(\ref{masterxx}), we get back to the original particle path integral $Z_{\rm particle}$ in eq.(\ref{particlez}). 
On the other hand, if we integrate out the field  $\tau_\mu$,  we find  the equation of motion, 
\be
(\tau_\mu+A_\mu)\phi_0^2=\frac{1}{2\pi}\epsilon^{\mu\nu\rho}\d_\nu \lambda_\rho\; .
\ee
By replacing this in eq.(\ref{masterxx}), we obtain the dual path integral, for the Lagrange multipliers $\lambda_\mu$, as in eq.(\ref{pvdual}).

We have then proven, at the level of partition functions, the particle-vortex duality or equivalence between eqs.(\ref{particlez}) and (\ref{pvdual}).
We will now use the results summarized in this section to prove the three-dimensional bosonization in eqs.(\ref{bosonization})-(\ref{papa}).

%%%%%%%%%%%%%%%%%%%%%%%%%%%%%%%%%%%%%%%%%%%%%%%%%%%%%%%%%%%%%%%%%%%%%%%%%%%%%%%%%%%%%%%%
\section{Proof of  the three-dimensional bosonization duality}\label{section3}
%%%%%%%%%%%%%%%%%%%%%%%%%%%%%%%%%%%%%%%%%%%%%%%%%%%%%%%%%%%%%%%%%%%%%%%%%%%%%%%%%%%%%%%%

In this section we provide a proof of the basic three-dimensional bosonization duality, in its mass deformed version, 
as written in eqs.(\ref{bosonization2})-(\ref{papa}).

%{\bf Proof of basic bosonization relation}
We first discuss the Burgess-Quevedo map (BQ-map). This should be thought of as a bosonization relation that can be derived in a self-consistent manner
\cite{Burgess:1993np,Burgess:1994tm,Fradkin:1994tt},  see \cite{Schaposnik:1995np} for a careful account of the BQ-map.
 We are interested in the formulation presented in the paper \cite{Burgess:1994tm}, that proceeds  by explicitly integrating out 
{\it massive } fermions {\em at low energies} (the energies are $E$ much smaller than the mass $m$) in the presence of a vector field background.
 In fact, we approximate the  fermionic determinant by  calculating 
 a fermion-loop with  only two external vector insertions. On top of this, we approximate this result for the case of large masses 
 (see the paper  \cite{Barci:1995iy} for details). Both approximations are well-justified in a $\frac{k}{m}$-expansion. After various 
 algebraic manipulations described 
 in  \cite{Schaposnik:1995np}, one obtains the  {approximate} relation, 
\be
Z_{\rm fermion}[S;m]=Z_{\rm gauge}[S]=\int {\cal D}\lambda_\mu e^{-i\left[\frac{1}{2k_3}\epsilon^{\mu\nu\rho}\lambda_\mu \d_\nu\lambda_\rho+
\epsilon^{\mu\nu\rho}\lambda_\mu\d_\nu S_\rho\right]}\;,
\ee
where $k_3=sign(m)/(4\pi)$\footnote{Note that \cite{Burgess:1994tm} has a different coefficient, but the original article \cite{Redlich:1983dv} quoted there 
implies this value. Indeed, integrating over $\lambda_\mu$ we get $S_{\rm eff}=\frac{k_3}{2}\int d^3x \epsilon^{\mu\nu\rho} S_\mu \d_\nu S_\rho$, and 
eq. (4.26) of the paper \cite{Redlich:1983dv} gives the coefficient as $k_3/2=sign(m) e^2/(8\pi)$. In this sense, the 
parameter $m$ is also present in $Z_{\rm gauge}[S]$.}.
Defining $\tilde \lambda_\mu=2\pi\lambda_\mu$, for $m>0$, this becomes\footnote{Note that in the work \cite{Barci:1995iy} a generic 
expression for the fermionic determinant and the BQ-map was obtained. The limit $m\rightarrow \infty$ of their expression, gives the same path integral quoted 
in eq.(\ref{bosoold}), after a  suitable rescaling in eq.(22) of   \cite{Barci:1995iy}.}
\be
Z_{\rm fermion}[S;m]=Z_{\rm gauge}[S]=\int {\cal D}\tilde \lambda_\mu e^{-2iS_{\rm CS}[\tilde \lambda]-iS_{\rm BF}[\tilde \lambda,S]}\;.
\label{bosoold}
\ee
It is nice to notice that we can supplement the BQ-map in eq.(\ref{bosoold}), extending  it to the situation  in which 
the system is in the presence of topological objects, like the singular vortices of the previous section.
Indeed, representing these vortices by a multiple-valued angle $\theta_{\rm vortex}$, the associated current $j^\mu_{\rm vortex}$  as  defined 
in eq.(\ref{yyyy}), and replacing $S_\mu\to S_\mu + \partial_\mu \theta_{\rm vortex}$ in the BQ-map of eq.(\ref{bosoold}), we find
\bea
& & Z_{\rm fermion+vortex}[S;m]= \int {\cal D}\psi{\cal D}\bar\psi e^{i\int \bar\psi(\dslash+m +S +\dslash \theta_{\rm vortex})\psi},\nonumber\\
& & Z_{\rm gauge+vortex}[S] =\int {\cal D}\tilde \lambda_\mu e^{-2iS_{\rm CS}[\tilde \lambda]-iS_{\rm BF}
[\tilde \lambda,S] -i\int d^3 x j^\mu_{\rm vortex} \tilde{\lambda}_\mu},\nonumber\\
& & Z_{\rm fermion+vortex}[S;m]= Z_{\rm gauge+vortex}[S] .\label{zaza}
\eea
The last equality in eq.(\ref{zaza}) is valid, as discussed above,  in the  regime
 of low energies (or large mass, in a $\frac{k}{m}$-expansion).
 %, justifying the approximate calculation of the fermionic determinant with only two  external insertions.

We will now use the BQ-map in the versions discussed above, together with the particle-vortex duality derived  in the previous section,
to prove the three dimensional bosonization duality in eqs.(\ref{bosonization})-(\ref{papa}).

We proceed as follows: first we set $\phi_0=$ constant  on both sides of particle-vortex duality, eqs.(\ref{particlez})-(\ref{pvdual}). 
This implies that the corresponding vortices are point-like and singular.
Then, we consider the situation in which we probe the system with very small energies, specifically
 $E\ll \phi_0^2$, so that we can neglect the Maxwell kinetic term
in comparison with the BF kinetic term in eq.(\ref{pvdual}). In this situation, the dynamics consists of 
point-like vortices coupled to a  Chern-Simons gauge field.

Now,  we add $S_{\rm CS}[A]+S_{\rm BF}[A;S]$ to the actions in both path integrals (which adds 'flux' to both sides) and integrate over $A_\mu$
as well. We obtain the equality of the modified particle path integrals for the two systems, one with particles and flux and the other with 
vortices and flux. On the particle with flux side
we have (note that we change $\lambda_\mu\rightarrow -\lambda_\mu$ in the path 
integral),
\be
Z'_{\rm particle+flux}[S]=\int{\cal D}A_\mu{\cal D}\theta e^{\left[iS_{\rm scalar}[\theta,A;\phi_0]+iS_{\rm CS}[A]+iS_{\rm BF}[A;S]\right]}
=Z_{\rm scalar+flux}[S]\;,\label{nana}
\ee
which as we can see is equal to the scalar+flux path integral in the bosonization relation of eqs.(\ref{bosonization})-(\ref{papa}). 
Notice that although $\phi_0$ appears 
here, it is not a true parameter. Indeed, by rescaling the dimensionless $\theta$ by $\phi_0$ we simply construct a scalar with the canonical dimension. In 
other words, $\phi_0$ simply defines units. 

On the other hand, on the vortex side of the duality, we are left
with a modified vortex path integral,
\be
Z'_{\rm vortex+flux}[S]=\int {\cal D}A_\mu{\cal D}\lambda_\mu e^{iS_{\rm BF}[\lambda;A]+iS_{\rm BF}[A;S]+iS_{\rm CS}[A]+i\int d^3x j^\mu_{\rm vortex}
\lambda_\mu}.\label{papap}
\ee
Evaluating the integral over $A_\mu$, we obtain the equation of motion,
\be
dA=-dS-d\lambda\;,
\ee
which when substituted back into the path integral  of eq.(\ref{papap}) gives,
\be
Z'_{\rm vortex+flux}[S]=\int {\cal D}\lambda_\mu e^{-iS_{\rm CS}[S]-iS_{\rm CS}[\lambda]-iS_{\rm BF}[S,\lambda] +i\int d^3x 
j^\mu_{\rm vortex}\lambda_\mu}.\label{zpvf}
\ee
Now, we redefine the path integral variable (with trivial Jacobian) by 
\be
%\lambda_\mu+S_\mu=\sqrt{2}\tilde\lambda_\mu +\frac{S_\mu}{\sqrt{2}}\Rightarrow 
\lambda_\mu=\sqrt{2}\tilde \lambda_\mu+ S_\mu\left(-1+\frac{1}{\sqrt{2}}\right)\;,
\label{cambio}\ee
to finally obtain
\be
Z'_{\rm vortex+flux}[S]=Z'_{\rm gauge+vortex}[S]e^{-\frac{i}{2}S_{\rm CS}[S]-i\int d^3x j^\mu_{\rm vortex}S_\mu\left(-1+\frac{1}{\sqrt{2}}\right)}\;.\label{tata}
\ee
Two comments are in order. First, we have identified the gauge path integral from the BQ-map  \cite{Burgess:1994tm} 
in the presence of non-trivial topology, with the explicit insertion of the vortex current, as in eq.(\ref{zaza}). Note however
that, since the vortex current multiplies $\lambda_\mu$, and not $\tilde \lambda_\mu$, we obtain an extra term coupling it to $S_\mu$, 
and also we get a $\sqrt{2}$ factor in the $j_\mu^v \tilde{\lambda}^\mu$ term of eq.(\ref{zaza}), which is why we have put a prime on $Z_{\rm gauge+vortex}$. 
This will translate into the same factors on the fermion-vortex side.  Second, related to  eq.(\ref{cambio}), we see that if the field 
$\lambda_\mu$ has quantized flux across a two sphere, the flux of $\tilde{\lambda}_\mu$  will not be quantized. This is a 
shortcoming of the change in eq.(\ref{cambio}). 

Then combining the various expressions in eqs.(\ref{zaza}),(\ref{nana}) and (\ref{tata}), we obtain a chain of equalities, 
\bea
Z_{\rm scalar+flux}[S]&=&Z'_{\rm particle+flux}[S]=Z'_{\rm vortex+flux}[S]\cr
&=&Z'_{\rm gauge+vortex}[S]e^{-\frac{i}{2}S_{\rm CS}[S]
-i\int d^3x j^\mu_{\rm vortex}S_\mu\left(-1+\frac{1}{\sqrt{2}}\right)}\cr
&=&Z'_{\rm fermion+vortex}[S]e^{-\frac{i}{2}S_{\rm CS}[S]-i\int d^3x j^\mu_{\rm vortex}S_\mu\left(-1+\frac{1}{\sqrt{2}}\right)}.\label{finalresult}
\eea
%Note that the insertion of the vortex current in (\ref{zpvf}) can be obtained by the replacement
%\be
%S_\mu\rightarrow S_\mu-\d_\mu\theta_{\rm vortex}\;,\label{Sreplace}
%\ee 
%where as before $\theta_{\rm vortex}$ contains the nontrivial topology sectors, so the same shift can be used to define the addition of the vortex 
%in $Z_{\rm gauge+vortex}$ and $Z_{\rm fermion+vortex}$. 

Focusing our attention on the first and last terms of the chain of equalities in eq.(\ref{finalresult}), we find that this 
is  the three-dimensional bosonization duality (\ref{bosonization2})-(\ref{papa}) that we wanted to prove (in the presence of non-trivial topology).

 Our derivation  needs the addition of the vortex coupling
$-\sqrt{2}\bar\psi\gamma^\mu\psi \d_\mu\theta_{\rm vortex}$ on the fermionic side---the term $\dslash \theta_{\rm vortex}$ eq.(\ref{zaza}), 
multiplied by $\sqrt{2}$. On the scalar side, we have an integration
over the full $\theta$ variable, which can be split into an integral over $\theta_{\rm smooth}$ and a sum over $\theta_{\rm 
vortex}$ sectors. Hence the presence of topology (nontrivial $\theta_{\rm vortex}$). Both the fermionic and bosonic  sides  of the 
three-dimensional bosonization duality 
are more general than initially assumed and were both supplemented by the presence of non-trivial topology. One may consider 
(after the derivation is complete) 
 the limit $\theta_{\rm vortex}=0$, to get back to the situation
with no topology in eqs.(\ref{bosonization2})-(\ref{papa}). This completes our derivation.
Once again, we emphasize  that the result has only been obtained for small energies $E\ll m=\phi_0^2$, $E\ll \a$.

%%%%%%%%%%%%%%%%%%%%%%%%%%%%%%%%%%%%%%%%%%%%%%%%%%%%%%%%%%%%%%%%%%%%%%%%%%%%%%%%%%%%%%%%
\section{Discussion and conclusions}\label{conclusion}
%%%%%%%%%%%%%%%%%%%%%%%%%%%%%%%%%%%%%%%%%%%%%%%%%%%%%%%%%%%%%%%%%%%%%%%%%%%%%%%%%%%%%%%%

We set out to prove the basic three-dimensional  bosonization relation in eq.(\ref{bosonization}), or its generalization by a mass in 
eqs.(\ref{bosonization2})-(\ref{papa}), which are at the basis of the derivation of the original web of dualities. Indeed, once the validity 
of eq.(\ref{bosonization}) is assumed, it can be used to prove Son's conjecture  \cite{Son:2015xqa}.
We have extended Son's relation to the case in which both fundamental and composite fermions are massive (with the same mass). 
This version of the conjecture
has been put on a firmer basis in our work.
% Hence,  the interesting equivalence between a massless Dirac fermion and a composite Dirac fermion describing the low energy limit of 
%the half-filled first Landau level of a Fermi liquid. 

We used a combination of the Burgess-Quevedo map, which is a Buscher-like correspondence between bosonic and fermionic theories 
\cite{Burgess:1994tm} and assumed the validity of the particle-vortex duality as defined in \cite{Murugan:2014sfa}. With this, we have shown that 
eqs.(\ref{bosonization})-(\ref{papa}) hold {\em at low energies} $E\ll m=\phi_0^2, E\ll \a$, with the addition of a vortex current term. 

The vortex current term would not be relevant for dualities between two bosonic theories, or between two fermionic ones, for these would
need to apply twice (in opposite directions) the basic bosonization duality of eq.(\ref{bosonization}). But it would influence other Bose to Fermi dualities,
for which we would apply it an odd number of times.

It is  significant  the fact that the derivation was only valid  for energies $E\ll m=\phi_0^2, E\ll \a$. It was already understood that 
the existence of dual pairs were valid for the  low energy
theories. Following our procedure,  we can only obtain eq.(\ref{bosonization}) for energies below $m$, which makes the $m\rightarrow 0$ limit 
singular. That means that results like the original Son's conjecture  \cite{Son:2015xqa}, are harder to understand, and would need extra  
arguments, to ensure their validity for  the $m\rightarrow0$ limit.

%%%%%%%%%%%%%%%%%%%%%%%%%%%%%%%%%%%%%%%%%%%%%%%%%%%%%%%%%%%%%%%%%%%%%%%%%%%%%%%%%%%%%%%%
%\section*{Note Added}
%%%%%%%%%%%%%%%%%%%%%%%%%%%%%%%%%%%%%%%%%%%%%%%%%%%%%%%%%%%%%%%%%%%%%%%%%%%%%%%%%%%%%%%%

%%%%%%%%%%%%%%%%%%%%%%%%%%%%%%%%%%%%%%%%%%%%%%%%%%%%%%%%%%%%%%%%%%%%%%%%%%%%%%%%%%%%%%%%
\section*{Acknowledgements}
%%%%%%%%%%%%%%%%%%%%%%%%%%%%%%%%%%%%%%%%%%%%%%%%%%%%%%%%%%%%%%%%%%%%%%%%%%%%%%%%%%%%%%%%

We wish to than S. Prem Kumar for discussions. We also want to thank also Jeff Murugan, Fidel Schaposnik and David Tong for useful comments on 
the manuscript. 
The work of HN is supported in part by CNPq grant 304006/2016-5 and FAPESP grant 2014/18634-9. HN would also 
like to thank the ICTP-SAIFR for their support through FAPESP grant 2016/01343-7. HN would also like to thank the Swansea
Physics Department for hospitality during the period when this project was started, and to the Royal Society for Newton Mobility Grant 
NI160034 to visit Swansea. CN is Wolfson Fellow of the Royal Society.

%%%%%%%%%%%%%%%%%%%%%%%%%%%%%%%%%%%%%%%%%%%%%%%%%%%%%%%%%%%%%%%%%%%%%%%%%%%%%%%%%%%%%%%%
\bibliography{dualityweb}

\providecommand{\href}[2]{#2}\begingroup\raggedright\begin{thebibliography}{10}

\bibitem{Son:2015xqa}
D.~T. Son, ``{Is the Composite Fermion a Dirac Particle?},''
  \href{http://dx.doi.org/10.1103/PhysRevX.5.031027}{{\em Phys. Rev.} {\bf X5}
  (2015) no.~3, 031027},
\href{http://arxiv.org/abs/1502.03446}{{\tt arXiv:1502.03446
  [cond-mat.mes-hall]}}.
%%CITATION = ARXIV:1502.03446;%%.

\bibitem{Murugan:2016zal}
J.~Murugan and H.~Nastase, ``{Particle-vortex duality in topological insulators
  and superconductors},''
\href{http://arxiv.org/abs/1606.01912}{{\tt arXiv:1606.01912 [hep-th]}}.
%%CITATION = ARXIV:1606.01912;%%.

\bibitem{Seiberg:2016gmd}
N.~Seiberg, T.~Senthil, C.~Wang, and E.~Witten, ``{A Duality Web in 2+1
  Dimensions and Condensed Matter Physics},''
  \href{http://dx.doi.org/10.1016/j.aop.2016.08.007}{{\em Annals Phys.} {\bf
  374} (2016)  395--433},
\href{http://arxiv.org/abs/1606.01989}{{\tt arXiv:1606.01989 [hep-th]}}.
%%CITATION = ARXIV:1606.01989;%%.

\bibitem{Karch:2016sxi}
A.~Karch and D.~Tong, ``{Particle-Vortex Duality from 3d Bosonization},''
  \href{http://dx.doi.org/10.1103/PhysRevX.6.031043}{{\em Phys. Rev.} {\bf X6}
  (2016) no.~3, 031043},
\href{http://arxiv.org/abs/1606.01893}{{\tt arXiv:1606.01893 [hep-th]}}.
%%CITATION = ARXIV:1606.01893;%%.

\bibitem{Metlitski:2015eka}
M.~A. Metlitski and A.~Vishwanath, ``{Particle-vortex duality of
  two-dimensional Dirac fermion from electric-magnetic duality of
  three-dimensional topological insulators},''
  \href{http://dx.doi.org/10.1103/PhysRevB.93.245151}{{\em Phys. Rev.} {\bf
  B93} (2016) no.~24, 245151},
\href{http://arxiv.org/abs/1505.05142}{{\tt arXiv:1505.05142
  [cond-mat.str-el]}}.
%%CITATION = ARXIV:1505.05142;%%.

\bibitem{Hsin:2016blu}
P.-S. Hsin and N.~Seiberg, ``{Level/rank Duality and Chern-Simons-Matter
  Theories},'' \href{http://dx.doi.org/10.1007/JHEP09(2016)095}{{\em JHEP} {\bf
  09} (2016)  095},
\href{http://arxiv.org/abs/1607.07457}{{\tt arXiv:1607.07457 [hep-th]}}.
%%CITATION = ARXIV:1607.07457;%%.

\bibitem{Karch:2016aux}
A.~Karch, B.~Robinson, and D.~Tong, ``{More Abelian Dualities in 2+1
  Dimensions},'' \href{http://dx.doi.org/10.1007/JHEP01(2017)017}{{\em JHEP}
  {\bf 01} (2017)  017},
\href{http://arxiv.org/abs/1609.04012}{{\tt arXiv:1609.04012 [hep-th]}}.
%%CITATION = ARXIV:1609.04012;%%.

\bibitem{Aharony:2016jvv}
O.~Aharony, F.~Benini, P.-S. Hsin, and N.~Seiberg, ``{Chern-Simons-matter
  dualities with $SO$ and $USp$ gauge groups},''
\href{http://arxiv.org/abs/1611.07874}{{\tt arXiv:1611.07874
  [cond-mat.str-el]}}.
%%CITATION = ARXIV:1611.07874;%%.

\bibitem{Kachru:2016aon}
S.~Kachru, M.~Mulligan, G.~Torroba, and H.~Wang, ``{Nonsupersymmetric dualities
  from mirror symmetry},''
  \href{http://dx.doi.org/10.1103/PhysRevLett.118.011602}{{\em Phys. Rev.
  Lett.} {\bf 118} (2017) no.~1, 011602},
\href{http://arxiv.org/abs/1609.02149}{{\tt arXiv:1609.02149 [hep-th]}}.
%%CITATION = ARXIV:1609.02149;%%.

\bibitem{Radicevic:2016wqn}
D.~Radicevic, D.~Tong, and C.~Turner, ``{Non-Abelian 3d Bosonization and
  Quantum Hall States},'' \href{http://dx.doi.org/10.1007/JHEP12(2016)067}{{\em
  JHEP} {\bf 12} (2016)  067},
\href{http://arxiv.org/abs/1608.04732}{{\tt arXiv:1608.04732 [hep-th]}}.
%%CITATION = ARXIV:1608.04732;%%.

\bibitem{Aharony:2015mjs}
O.~Aharony, ``{Baryons, monopoles and dualities in Chern-Simons-matter
  theories},'' \href{http://dx.doi.org/10.1007/JHEP02(2016)093}{{\em JHEP} {\bf
  02} (2016)  093},
\href{http://arxiv.org/abs/1512.00161}{{\tt arXiv:1512.00161 [hep-th]}}.
%%CITATION = ARXIV:1512.00161;%%.

\bibitem{Burgess:1993np}
C.~P. Burgess and F.~Quevedo, ``{Bosonization as duality},''
  \href{http://dx.doi.org/10.1016/0550-3213(94)90332-8}{{\em Nucl. Phys.} {\bf
  B421} (1994)  373--390},
\href{http://arxiv.org/abs/hep-th/9401105}{{\tt arXiv:hep-th/9401105
  [hep-th]}}.
%%CITATION = HEP-TH/9401105;%%.

\bibitem{Burgess:1994tm}
C.~P. Burgess, C.~A. Lutken, and F.~Quevedo, ``{Bosonization in higher
  dimensions},'' \href{http://dx.doi.org/10.1016/0370-2693(94)00963-5}{{\em
  Phys. Lett.} {\bf B336} (1994)  18--24},
\href{http://arxiv.org/abs/hep-th/9407078}{{\tt arXiv:hep-th/9407078
  [hep-th]}}.
%%CITATION = HEP-TH/9407078;%%.

\bibitem{Fradkin:1994tt}
E.~H. Fradkin and F.~A. Schaposnik, ``{The Fermion - boson mapping in
  three-dimensional quantum field theory},''
  \href{http://dx.doi.org/10.1016/0370-2693(94)91374-9}{{\em Phys. Lett.} {\bf
  B338} (1994)  253--258},
\href{http://arxiv.org/abs/hep-th/9407182}{{\tt arXiv:hep-th/9407182
  [hep-th]}}.
%%CITATION = HEP-TH/9407182;%%.

\bibitem{Schaposnik:1995np}
F.~A. Schaposnik, ``{A Comment on bosonization in d $\geq$ two-dimensions},''
  \href{http://dx.doi.org/10.1016/0370-2693(95)00776-H}{{\em Phys. Lett.} {\bf
  B356} (1995)  39--44},
\href{http://arxiv.org/abs/hep-th/9505049}{{\tt arXiv:hep-th/9505049
  [hep-th]}}.
%%CITATION = HEP-TH/9505049;%%.

\bibitem{Filothodoros:2016txa}
E.~G. Filothodoros, A.~C. Petkou, and N.~D. Vlachos, ``{$3d$ fermion-boson map
  with imaginary chemical potential},''
\href{http://arxiv.org/abs/1608.07795}{{\tt arXiv:1608.07795 [hep-th]}}.
%%CITATION = ARXIV:1608.07795;%%.

\bibitem{Murugan:2014sfa}
J.~Murugan, H.~Nastase, N.~Rughoonauth, and J.~P. Shock, ``{Particle-vortex and
  Maxwell duality in the $AdS_4\times \mathbb{CP}^3$/ABJM correspondence},''
  \href{http://dx.doi.org/10.1007/JHEP10(2014)051}{{\em JHEP} {\bf 10} (2014)
  51},
\href{http://arxiv.org/abs/1404.5926}{{\tt arXiv:1404.5926 [hep-th]}}.
%%CITATION = ARXIV:1404.5926;%%.

\bibitem{Burgess:2000kj}
C.~P. Burgess and B.~P. Dolan, ``{Particle vortex duality and the modular
  group: Applications to the quantum Hall effect and other 2-D systems},''
  \href{http://dx.doi.org/10.1103/PhysRevB.63.155309}{{\em Phys. Rev.} {\bf
  B63} (2001)  155309},
\href{http://arxiv.org/abs/hep-th/0010246}{{\tt arXiv:hep-th/0010246
  [hep-th]}}.
%%CITATION = HEP-TH/0010246;%%.

\bibitem{Ramos:2005yy}
R.~O. Ramos, J.~F. Medeiros~Neto, D.~G. Barci, and C.~A. Linhares, ``{Abelian
  Higgs model effective potential in the presence of vortices},''
  \href{http://dx.doi.org/10.1103/PhysRevD.72.103524}{{\em Phys. Rev.} {\bf
  D72} (2005)  103524},
\href{http://arxiv.org/abs/hep-th/0506052}{{\tt arXiv:hep-th/0506052
  [hep-th]}}.
%%CITATION = HEP-TH/0506052;%%.

\bibitem{Ramos:2007hk}
R.~O. Ramos and J.~F. Medeiros~Neto, ``{Transition Point for Vortex
  Condensation in the Chern-Simons Higgs Model},''
  \href{http://dx.doi.org/10.1016/j.physletb.2008.07.097}{{\em Phys. Lett.}
  {\bf B666} (2008)  496--501},
\href{http://arxiv.org/abs/0711.0798}{{\tt arXiv:0711.0798 [hep-th]}}.
%%CITATION = ARXIV:0711.0798;%%.

\bibitem{Zee:2003mt}
A.~Zee, {\em {Quantum field theory in a nutshell}}.
\newblock
2003.
\newblock
%%CITATION = ISBN-9780691140346;%%.

\bibitem{Dasgupta:1981zz}
C.~Dasgupta and B.~I. Halperin, ``{Phase Transition in a Lattice Model of
  Superconductivity},''
\href{http://dx.doi.org/10.1103/PhysRevLett.47.1556}{{\em Phys. Rev. Lett.}
  {\bf 47} (1981)  1556--1560}.
%%CITATION = PRLTA,47,1556;%%.

\bibitem{Marino:1987tk}
E.~C. Marino, ``{Quantum Theory of Nonlocal Vortex Fields},''
\href{http://dx.doi.org/10.1103/PhysRevD.38.3194}{{\em Phys. Rev.} {\bf D38}
  (1988)  3194}.
%%CITATION = PHRVA,D38,3194;%%.

\bibitem{Marino:1992uu}
E.~C. Marino, ``{Duality, quantum vortices and anyons in
  Maxwell-Chern-Simons-Higgs theories},''
  \href{http://dx.doi.org/10.1006/aphy.1993.1046}{{\em Annals Phys.} {\bf 224}
  (1993)  225--274},
\href{http://arxiv.org/abs/hep-th/9208062}{{\tt arXiv:hep-th/9208062
  [hep-th]}}.
%%CITATION = HEP-TH/9208062;%%.

\bibitem{Lee:1989fw}
D.~H. Lee and M.~P.~A. Fisher, ``{Anyon superconductivity and the fractional
  quantum Hall effect},''
\href{http://dx.doi.org/10.1103/PhysRevLett.63.903}{{\em Phys. Rev. Lett.} {\bf
  63} (1989)  903--906}.
%%CITATION = PRLTA,63,903;%%.

\bibitem{Barci:1995iy}
D.~G. Barci, C.~D. Fosco, and L.~E. Oxman, ``{On bosonization in
  three-dimensions},''
  \href{http://dx.doi.org/10.1016/0370-2693(96)00224-9}{{\em Phys. Lett.} {\bf
  B375} (1996)  267--272},
\href{http://arxiv.org/abs/hep-th/9508075}{{\tt arXiv:hep-th/9508075
  [hep-th]}}.
%%CITATION = HEP-TH/9508075;%%.

\bibitem{Redlich:1983dv}
A.~N. Redlich, ``{Parity Violation and Gauge Noninvariance of the Effective
  Gauge Field Action in Three-Dimensions},''
\href{http://dx.doi.org/10.1103/PhysRevD.29.2366}{{\em Phys. Rev.} {\bf D29}
  (1984)  2366--2374}.
%%CITATION = PHRVA,D29,2366;%%.

\end{thebibliography}\endgroup
\bibliographystyle{utphys}
%%%%%%%%%%%%%%%%%%%%%%%%%%%%%%%%%%%%%%%%%%%%%%%%%%%%%%%%%%%%%%%%%%%%%%%%%%%%%%%%%%%%%%%%

\end{document}